\documentclass[ reprint,
superscriptaddress,
 amsmath,amssymb,
 aps, floatfix
]{revtex4-2}
\usepackage{graphicx}
\usepackage{dcolumn}
\usepackage{bm}
\usepackage{color}
\usepackage{amsmath}
\usepackage{braket}
\usepackage{amsfonts}
\usepackage{soul}
\usepackage{url}
\usepackage{array,tabularx}
\usepackage{graphics}
\usepackage{times}
\usepackage{subfigure}
\usepackage[export]{adjustbox}
\newcommand\x{3.5}

\begin{document}

\title{Gravitational lensing and tunneling of mechanical waves in synthetic curved spacetime}
\author{Sayan Jana}\email{sayanjana@tauex.tau.ac.il}
\affiliation{School of Mechanical Engineering, Tel Aviv University, Tel Aviv 69978, Israel}
\author{Lea Sirota}\email{leabeilkin@tauex.tau.ac.il}
\affiliation{School of Mechanical Engineering, Tel Aviv University, Tel Aviv 69978, Israel}

\begin{abstract}

Black holes are considered among the most fascinating objects that exist in our universe, since in the classical formalism nothing, even no light, can escape from their vicinity due to gravity. 
The gravitational potential causes the light to bend towards the hole, which is known by gravitational lensing.
Here we present a synthetic realization of this phenomenon in a lab-scale two-dimensional network of mechanical circuits, based on analogous condensed matter formalism of Weyl semimetals with inhomogeneous nodal tilt profiles.
Some of the underlying network couplings turn out as unstable and non-reciprocal, and are implemented by embedded active feedback interactions in an overall stabilized structure.
We demonstrate the lensing by propagating mechanical wavepackets through the network with a programmed funnel-like potential, achieving wave bending towards the circle center. 
We then demonstrate the versatility of our platform by reprogramming it to mimic quantum tunneling of particles through the event horizon, known by Hawking radiation, achieving an exceptional correspondence to the original mass loss rate within the hole. 
The network couplings and the potential can be further reprogrammed to realize other curvatures and associated relativistic phenomena.

\end{abstract}

\maketitle

In our universe, the existence of black holes \cite{frolov2011introduction} is theoretically predicted by Einstein's theory of general relativity through the spacetime singularity in the Schwarzschild metric \cite{misner1973gravitation}. This metric determines the curvature of spacetime geometry for which gravity is the outcome. In classical gravity event horizon is thought as a boundary between the black hole and the visible universe. 
Any object that crosses this boundary is dragged with the speed of light towards the center and ultimately becomes invisible. 
The resulting propagation of relativistic light in the curved spacetime around the hole is characterized by deflection, or lensing inward due to the gravitational potential.

Recently, a condensed matter analogue of the spacetime curvature has been obtained through Weyl semimetals (WSM) that are characterized by the Hamiltonian $H_{WSM}=\left(V_{t}+v_{f}\sigma\right)\cdot\textbf{k}$ with a spatially-varying tilt profile $V_{t}(r)$ \cite{volovik2016black,zubkov2018black,huang2018black,liang2019curved,kedem2020black,konye2022horizon}. 
Here, $r$ and $\textbf{k}$ are respectively the spatial coordinate and the momentum in three dimensions, spanned by Pauli matrices $\sigma$. 
The analogue was drawn pictorially by mapping Weyl cones into space-time geodesices (light cones), obtained from the Painleve \cite{painleve1921mecanique}-Gullstrand \cite{gullstrand1922allgemeine}-Lema$\hat{i}$tre \cite{lemaitre1933univers} coordinate system,
  \begin{equation}
    ds^{2}=c^2 dt^2-(dr-V(r)dt)^{2},
    \label{eq:eq2}
 \end{equation}
in which the spatial profile $V(r)$ incorporates the information of gravity, and the event horizon is denoted by the radius where $V(r)$ equals the speed of light $c$. 
In the WSM equivalence, the tilt $V_{t}(r)$ and Fermi velocity $v_{f}$ are respectively mapped to $V(r)$ and $c$ in \eqref{eq:eq2}, yielding the horizon along $r$ that satisfies $V_{t}(r)=v_{f}$. The resulting system can be represented by the quantum Bloch Hamiltonian \cite{konye2022anisotropic,haller2022black}
\begin{equation} \label{eq:gl_H}
\begin{split}
    H(k)&=\sum\nolimits_{j=x,y} t_j\left(\sigma_j-V_j\sigma_0\right)\sin{k_ja}\\&+t_z\sigma_z\left(2-\cos{k_xa}-\cos{k_ya}\right), 
    \end{split}
\end{equation} 
where $t_x,t_y$ indicate the spin-orbit coupling strength, $t_z$ is the nearest neighbour hopping parameter, $V_x,V_y$ are the $x,y$ projections of the tilting potential $V_t(r)$, and $a$ is the lattice constant.  
This mapping between two completely different fields, relativistic physics and quantum condensed matter, is valid also for other high-energy phenomena, such as Klein tunneling in graphene \cite{katsnelson2006chiral}, transport in solids \cite{bermond2022anomalous}, and chiral
anomaly in WSM \cite{behrends2019landau}. 
However, laboratory imitation of the high-energy effects, even via the condensed matter analogy \cite{jiang2020direct,sirota2022klein}, is not immediate. 

\begin{figure*}[htpb]
\begin{center}
    \begin{tabular}{c c c c}
    \textbf{(a)} & \textbf{(b)} & \textbf{(c)} & \textbf{(d)} \\ 
     & & & \\
     \includegraphics[height=5.3 cm, valign=c]{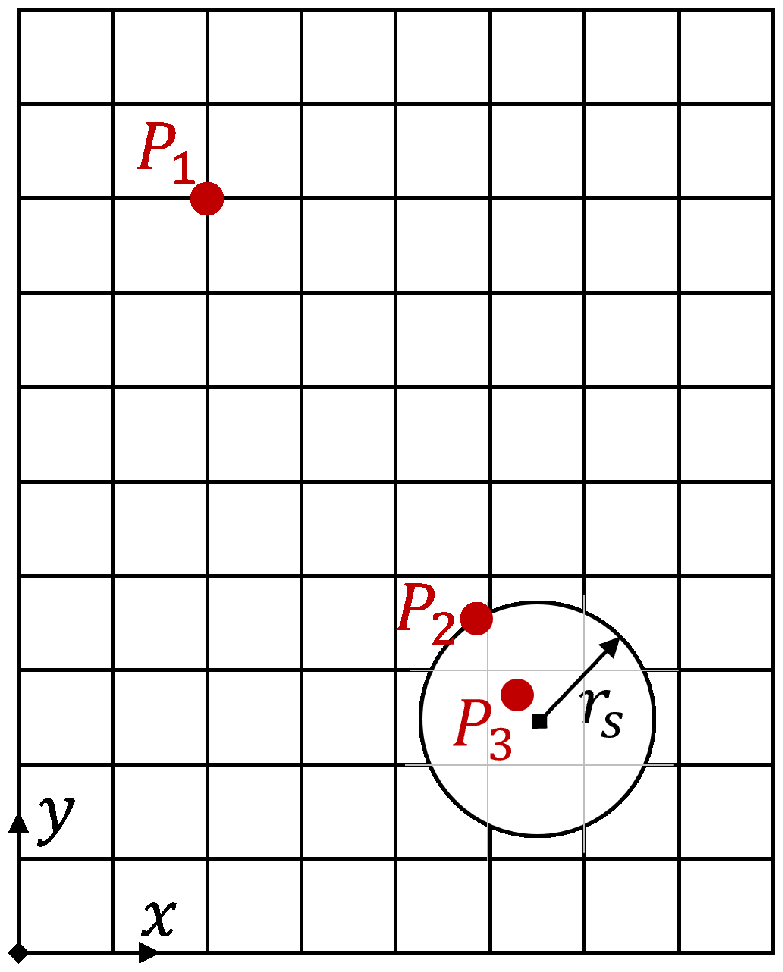}    &  \includegraphics[height=5.3 cm, valign=c]{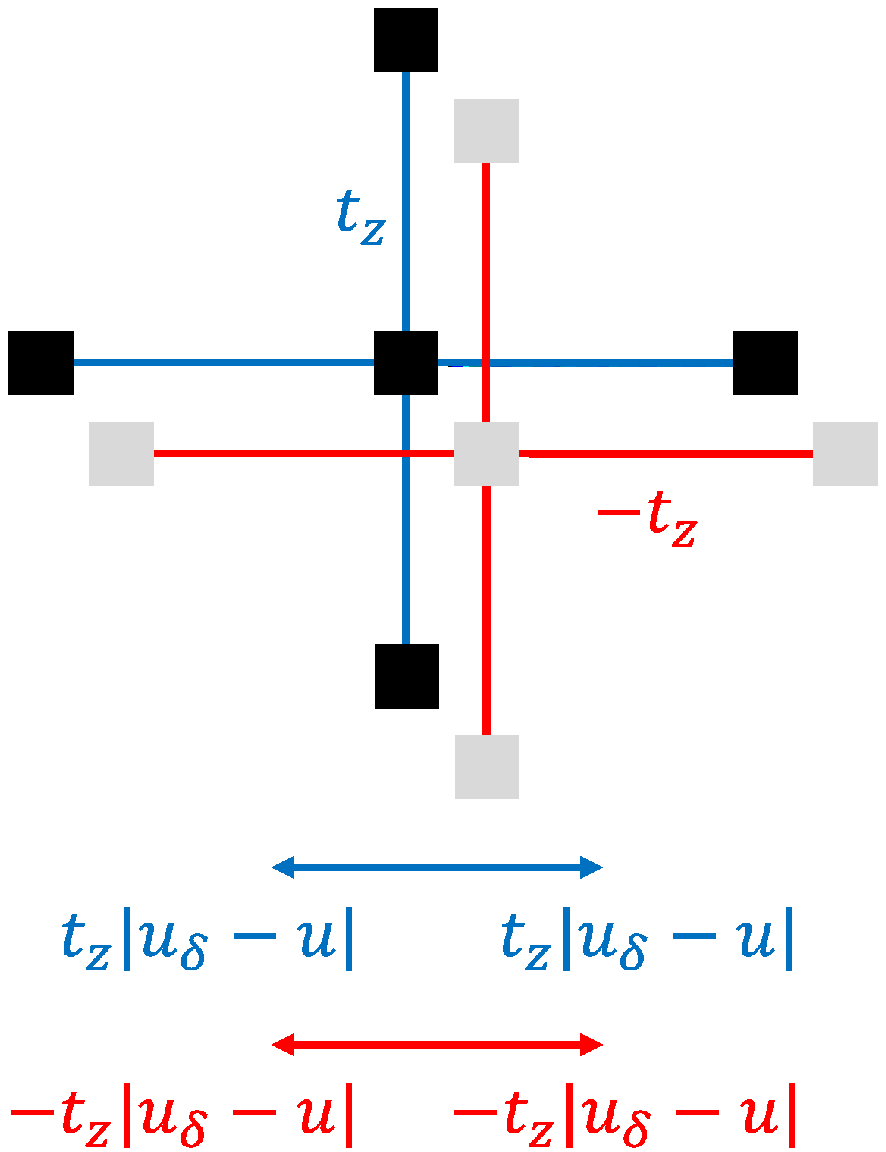}    & \includegraphics[height=5.3 cm, valign=c]{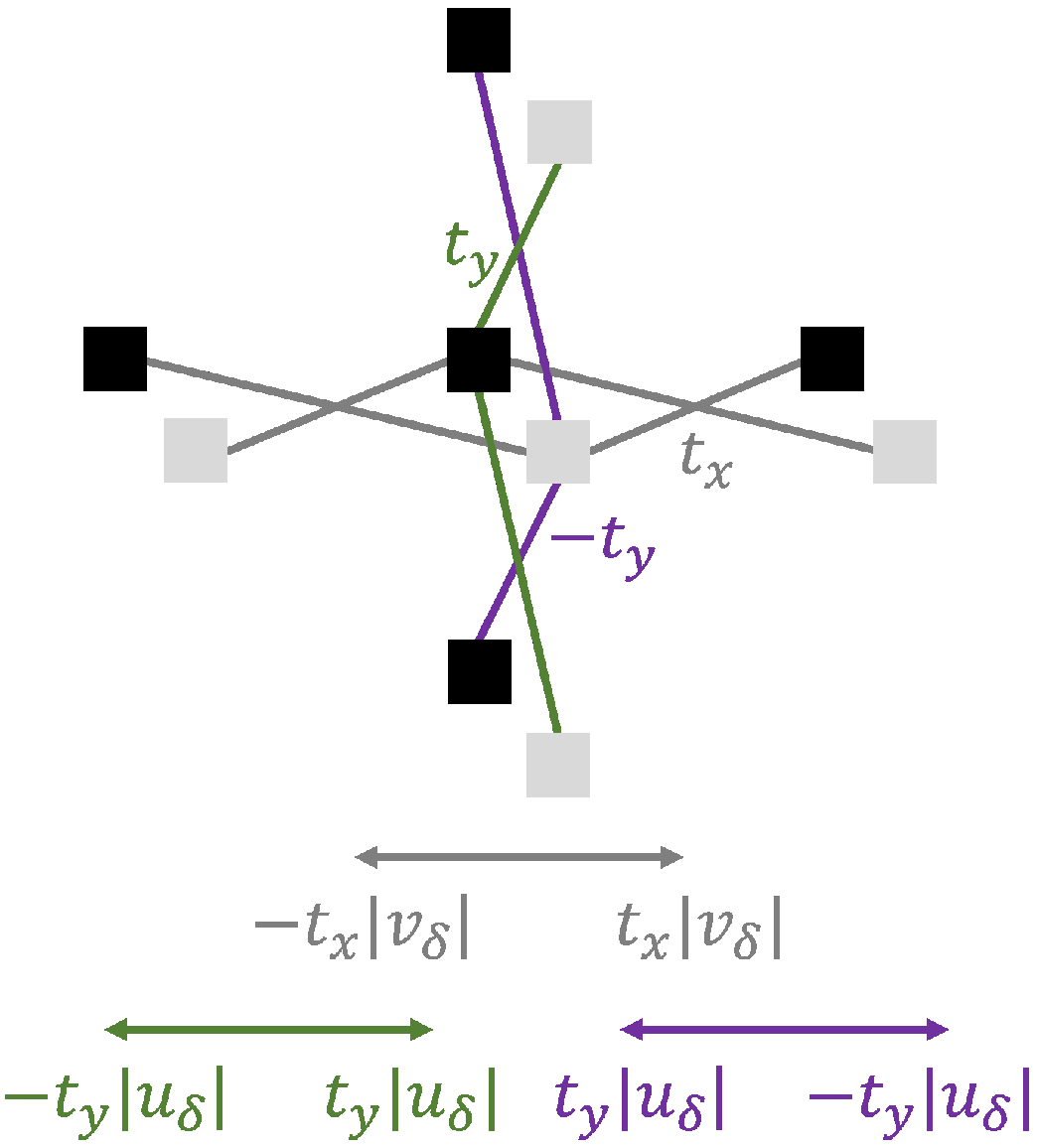}     &   \includegraphics[height=5.3 cm, valign=c]{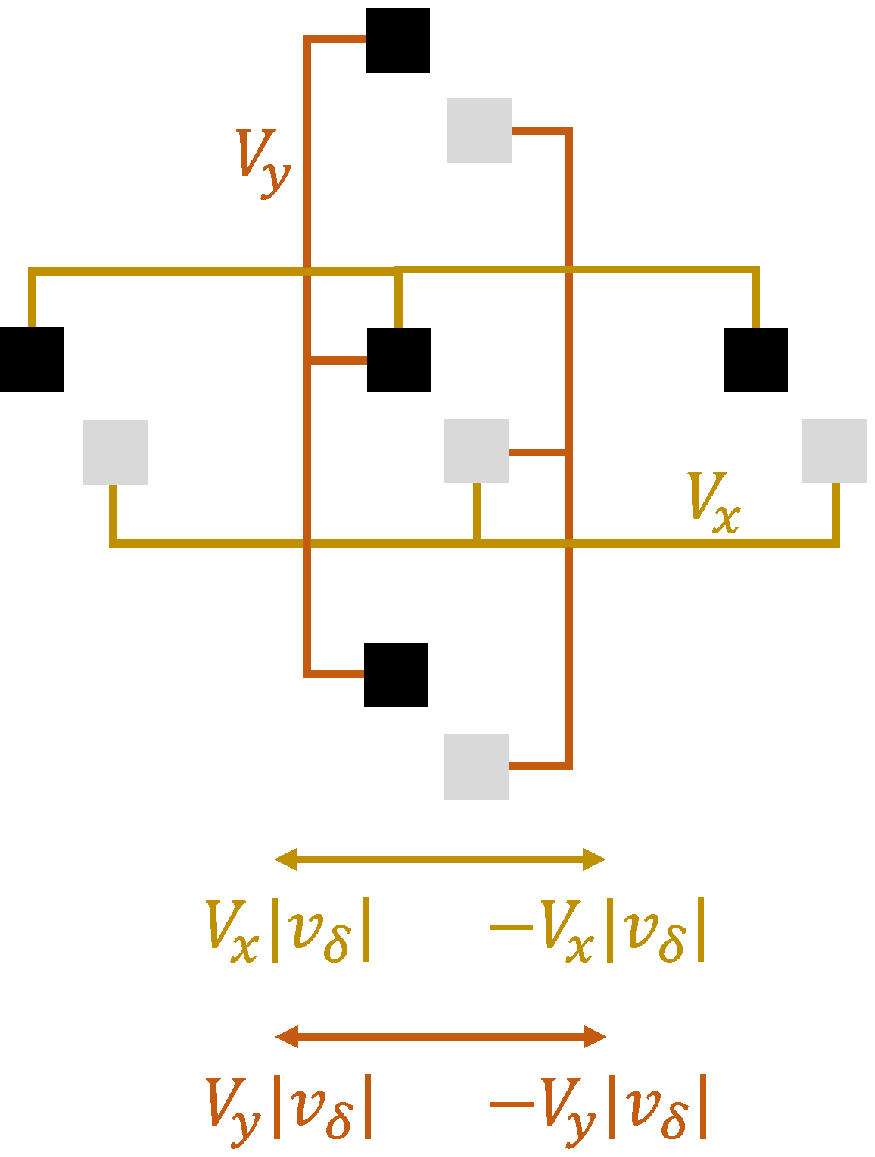} \\ & & &
    \end{tabular} 
\end{center}
\caption{The mechanical circuits network imitating curved spacetime. (a) The network schematic with a black hole of radius $r_s$. $P_{1,2,3}$ indicate representative locations outside the hole, on the horizon and inside the hole. (b),(c),(d) The unit cell with the target closed-loop couplings $+t_z$, $-t_z$, $it_x$, $t_y$, $-t_y$, $iV_x(x,y)$ and $iV_y(x,y)$, depicted in separate panels for clarity. Below the panels are plotted the forces that each element exerts on the masses, as implied by the Hamiltonian in \eqref{eq:gl_H}. $\delta$ indicates the nearest neighbor mass, $u$ is displacement and $v$ is velocity.}
\label{fig:model}
\end{figure*}
Here we present a framework for constructing an experimentally ready purely classical model, consisting of a network of active mechanical circuits \cite{susstrunk2015observation,nash2015topological,mousavi2015topologically,pal2017edge,chaunsali2018subwavelength,zhou2018quantum}, which realizes, based on the WSM formalism, a synthetic curved spacetime and the associated gravitational lensing.
In addition, our platform realizes another aspect of the black hole, which is Hawking radiation, also known as horizon tunneling. The latter stems from the astonishing discovery that in the quantum realm the phenomenon of completely black is not entirely true. 

Rather, it was shown that a black hole radiates \cite{hawking1974black,hawking1975particle,bekenstein2020black}, and this radiation exists as fluctuation of quantum fields near the horizon with temperature $T_{H}$, famously known as the Hawking temperature. Equivalently, Hawking radiation can be visualized as semi-classical tunneling of particles through the event horizon \cite{srinivasan1999particle,parikh2000hawking} by considering the following scenario: a pair of positive and negative energy particles is created  either just inside the horizon, where a positive energy particle tunnels outward, or just outside the event horizon, where a negative energy particle tunnels inward.
Tunneling of a particle with energy $E$ is compensated by an equivalent mass loss within the black hole for which the emission rate is given by
\begin{equation} \label{eq:eq1}
\Gamma_{H} \approx e^{-\frac{\hbar E}{K_{B}T_{H}}}.
\end{equation}  
Remarkably, Hawking temperature is proportional to the gravitational field strength $g$ as $K_{B}T_{H}$=$\hbar g/2\pi c$=$\hbar c^{3}/8\pi G M$, where $K_{B}$ is Boltzman constant, $\hbar$ is Planck constant, and $G$ stands for the universal gravitational constant. $T_{H}$ is inversely proportional to the black hole mass $M$, and for a small black hole (solar mass) currently present in our universe $T_{H}$ is of the order of $\approx 10^{-6}$ Kelvin \cite{robertson2012theory}. 
As this value is six orders of magnitude smaller than the cosmic microwave background, it is overshadowed, thus remaining undetected till now. This motivated a search for event horizon analogues, based on wave propagation in counter-flowing fluids, and associated ideas in photonics and Bose-Einstein condensates 
\cite{unruh1981experimental,garay2000sonic,leonhardt2000relativistic,leonhardt2002laboratory,giovanazzi2005hawking,schutzhold2005hawking,weinfurtner2011measurement,hu2019quantum,carot1999spherically,tajik2023experimental}.
In the proposed mechanical circuits platform we realize horizon tunneling based on the same WSM formalism as for gravitational lensing, solely by reprogramming the artificial potential in \eqref{eq:gl_H}. Our model  relies on the mechanical wave packet tunneling through synthetic horizon, which makes it completely distinct from pair production mechanism \cite{srinivasan1999particle,parikh2000hawking}.

The platform is illustrated in Fig. \ref{fig:model}(a). This is a two-dimensional square network with the black hole represented by a circle of radius $r_s$. 
This network needs to realize a synthetic curved spacetime, and thus its dynamical matrix needs to fully retrieve the quantum Hamiltonian in \eqref{eq:gl_H}.
The unit cell of the network, as depicted in Fig. \ref{fig:model}(b), (c), (d), consists of masses at two sites, $A$ (black squares), and $B$ (grey squares), with a single degree of freedom per site, e.g. vibrating out of plane with a displacement $u$ and velocity $v$. 
Following \eqref{eq:gl_H}, the masses are directly coupled by $t_z$ (blue bars), $-t_z$ (red bars), $iV_x(x,y)$ (yellow bars) and $iV_y(x,y)$ (orange bars), and are cross-coupled by $it_x$ (grey bars), $t_y$ (green bars) and $-t_y$ (violet bars). 
For the circle perimeter to represent the horizon, the quantum potential $V_t(r)$, and thus the corresponding couplings in the network, need to equal $v_f$ (or 1 in a normalized formulation) at the perimeter $r=r_s$ \cite{konye2022anisotropic,haller2022black}. To support the lensing phenomenon the potential strength also needs to increase toward the circle origin $r=0$. A specific expression is set at the model validation stage.

The nature of each coupling is illustrated below the panels with the subscript $\delta$ indicating the nearest connected mass. The $+t_z$ coupling is reciprocal and stable, and can be realized by a linear spring. The $-t_z$ coupling is reciprocal but is inherently unstable due to the negative sign, equivalently to a spring that expands when stretched.
The complex-valued couplings $it_x$, $iV_x(x,y)$ and $iV_y(x,y)$, relate to the masses velocities, as the Fourier transform of $v=\Dot{u}$ equals $i\Omega u$. These couplings, as well as $\pm t_y$, lack a restoring force and have an opposite sign at each end, being thus both non-reciprocal and unstable (but the total system is Hermitian). 

\begin{figure}[tb]
    \centering
    \begin{tabular}{c}
    \textbf{(a)}  \\
           \\
      \includegraphics[width=5.9 cm, valign=b]{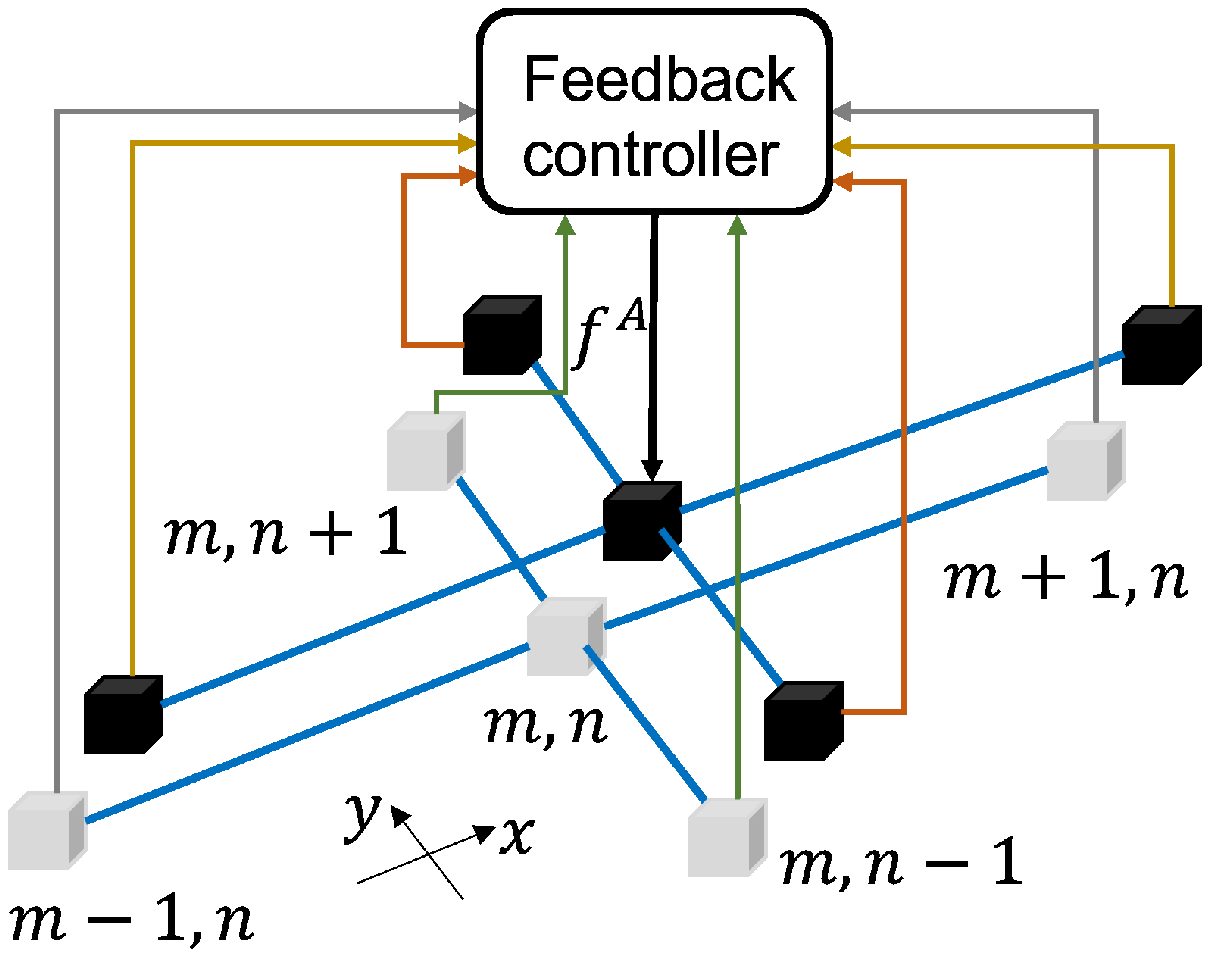}    \\ \textbf{(b)} \\  \\ \includegraphics[width=5.9 cm, valign=b]{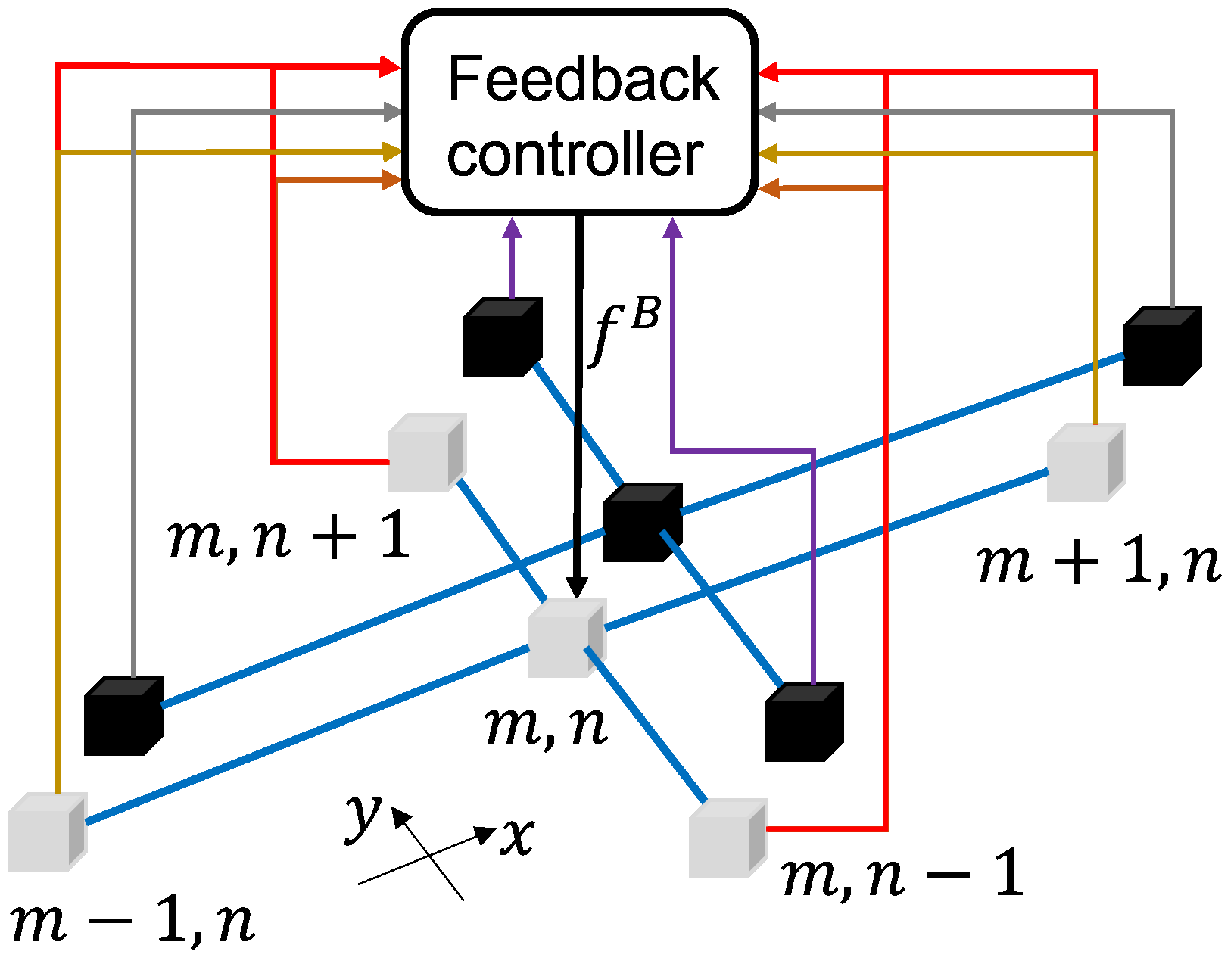}
    \end{tabular}
    \caption{Control mechanism implementation scheme. (a),(b) The actuation (black) and measurement (color) signals at the $\{m,n\}$ unit cell for the $A$ and $B$ sites, respectively.}
    \label{fig:control}
\end{figure}

As all the couplings except $+t_z$ cannot be implemented by passive elements, we implement them in our network using active feedback mechanism \cite{hofmann2019chiral,brandenbourger2019non,sirota2019tunable,darabi2020experimental,scheibner2020non,rosa2020dynamics,helbig2020generalized,sirota2020non,sirota2021real,sirota2021quantum,zhang2021acoustic}. This mechanism is embedded in a stable host network consisting of the $+t_z$ couplings only, shown by the blue bars in the schematic of Fig. \ref{fig:control}(a,b).
An active controller generates commands for external forces $f^{A},f^{B}$ at each unit cell, operating in a real-time closed loop. 
These forces are responsible for creating the couplings $-t_z$, $it_x$, $\pm t_y$, $iV_x(x,y)$ and $iV_y(x,y)$, as well as to stabilize the overall system.  
In the $\{m,n\}$ $A/B$ sites, Fig. \ref{fig:control}(a)/(b), the action of $f^{A/B}$ is based on velocity measurements of the $\{m\pm 1,n\}$ and $\{m,n\pm 1\}$ $A/B$ sites, and of the $\{m\pm 1,n\}$ $B/A$ sites, respectively indicated by the yellow, orange, and grey arrows, as well as displacement measurements of the $\{m,n\pm 1\}$ $B/A$ sites, indicated by the green/violet arrows. 
The action of $f^B$ alone is also based on displacement measurements of the $\{m\pm 1,n\}$ and $\{m,n\pm 1\}$ $B$ sites, indicated by the red arrows.
In addition, the action of $f^{A/B}$ is based on displacement measurements of $\{m,n\}$ $A/B$ sites (not indicated in the figure).  
The measured signals are fed back in real-time into the electronic controller, the gains of which are programmed into the matrix $C$. The algorithm is given by
\begin{equation} \label{eq:control_eq}
\begin{split}
&\left[\begin{array}{@{\mkern2mu} c @{\mkern2mu}}
    f^A \\
     f^B 
\end{array}\right]=-C\left[\begin{array}{@{\mkern2mu} c @{\mkern2mu}}
   \textbf{u}   \\
     \textbf{v}
\end{array}\right], \quad \def\arraystretch{1.5} C=\frac{1}{2}\left[\begin{array}{@{\mkern2mu} cc @{\mkern2mu}}
   C^A_u   & C^A_v \\
     C^B_u &  C^B_v 
\end{array}\right],  \\ &\begin{cases} \setlength{\arraycolsep}{1.8pt}  C^A_u=\left[\begin{array}{@{\mkern2mu} cccccccc @{\mkern2mu}}
     -\beta &
     0  &
     0 &
     0 &
     -t_y  &
     t_y &
     0 & 0
\end{array}\right] \\ \setlength{\arraycolsep}{1.8pt}  C^B_u=\left[\begin{array}{@{\mkern2mu} cccccccc @{\mkern2mu}}
     0 &
     t_y  &
     -t_y & 
     -(\beta+8t_z) &
     2t_z  &
     2t_z &
     2t_z & 2t_z
\end{array}\right] \\ \setlength{\arraycolsep}{1.8pt}
C^A_v=\left[\begin{array}{@{\mkern2mu} cccccccc @{\mkern2mu}}
     V_x &
     -V_x  &
     V_y  &
     -V_y &
     -t_x &
     t_x  &
     0  &
     0
\end{array}\right] \\ \setlength{\arraycolsep}{1.8pt} C^B_v=\left[\begin{array}{@{\mkern2mu} cccccccc @{\mkern2mu}}
     -t_x &
     t_x  &
     0  &
     0 &
     V_x &
     -V_x  &
     V_y  &
     -V_y
\end{array}\right] \end{cases}
\end{split}
\end{equation}
where 
$\setlength{\arraycolsep}{1.8pt} \textbf{u}^{A/B}=\left[\begin{array}{@{\mkern2mu} ccccc @{\mkern2mu}}u^{A/B}_{m,n} & u^{A/B}_{m,n+1} & u^{A/B}_{m,n-1} & u^B_{m+1,n} & u^B_{m-1,n} \end{array} \right]$ and $\setlength{\arraycolsep}{1.8pt} \textbf{v}^{A/B}=\left[\begin{array}{@{\mkern2mu} cccc @{\mkern2mu}}v^{A/B}_{m+1,n} & v^{A/B}_{m-1,n} & v^{A/B}_{m,n+1} & v^{A/B}_{m,n-1} \end{array}\right]$ constitute the measurement signals $\setlength{\arraycolsep}{1.8pt} \textbf{u}=[\begin{array}{@{\mkern2mu} cc @{\mkern2mu}}
     \textbf{u}^A & \textbf{u}^B
\end{array}]'$ and $\setlength{\arraycolsep}{1.8pt} \textbf{v}=[\begin{array}{@{\mkern2mu} cc @{\mkern2mu}}
     \textbf{v}^A & \textbf{v}^B
\end{array}]'$ for the $\{m,n\}$ unit cell, $V_{x/y}=V_{x/y}(x_{m,n},y_{m,n})$ is the potential,
and $\beta=-8t_z$ guarantees the overall system stability.

\begin{figure}[tb]
\begin{center}
\setlength{\tabcolsep}{-2pt}
    \begin{tabular}{c c}
    \textbf{(a)} & \textbf{(b)} $P_1$ \\
     \includegraphics[height=4.5 cm, valign=c]{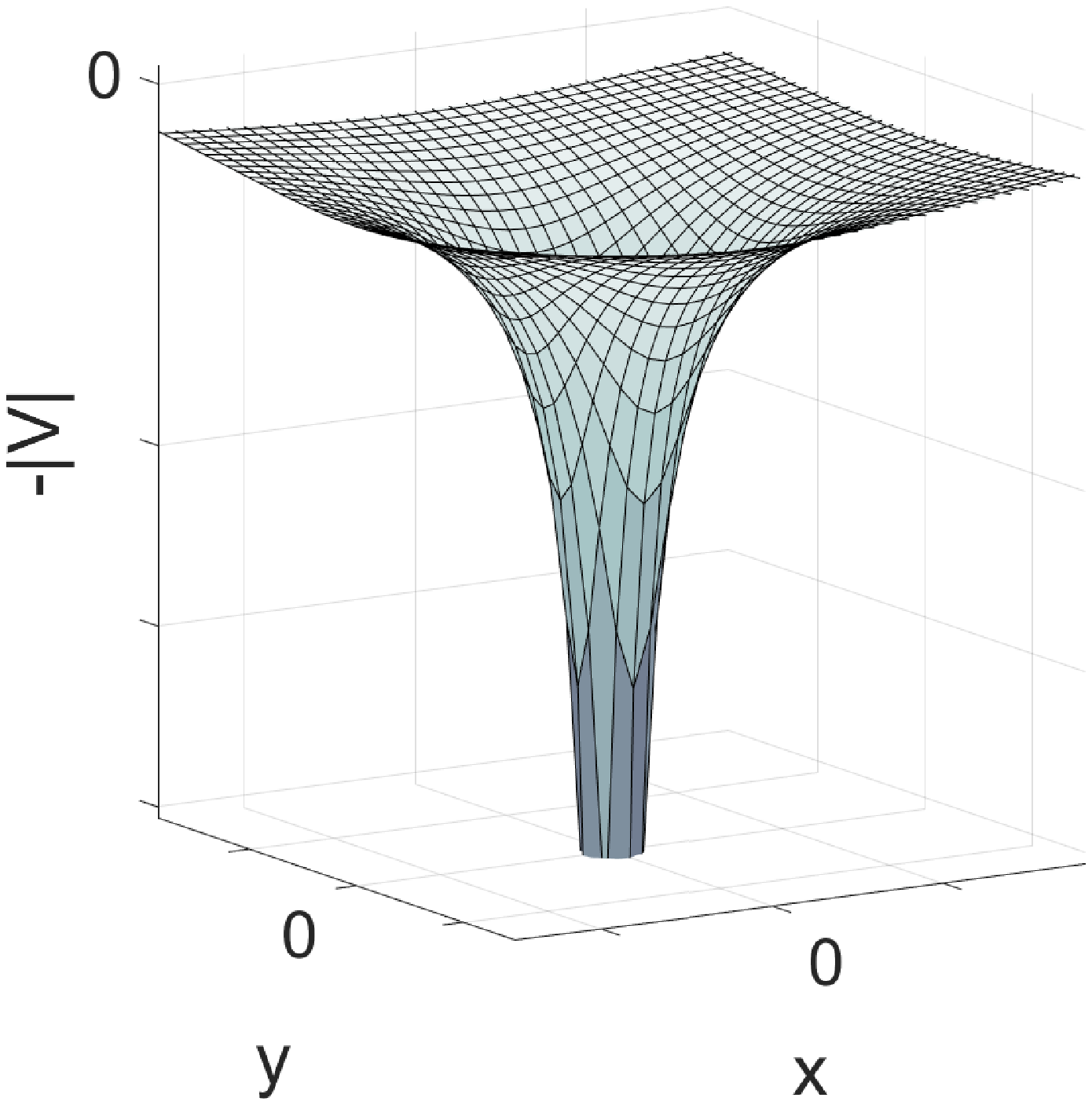}    &  \includegraphics[height=4.5 cm, valign=c]{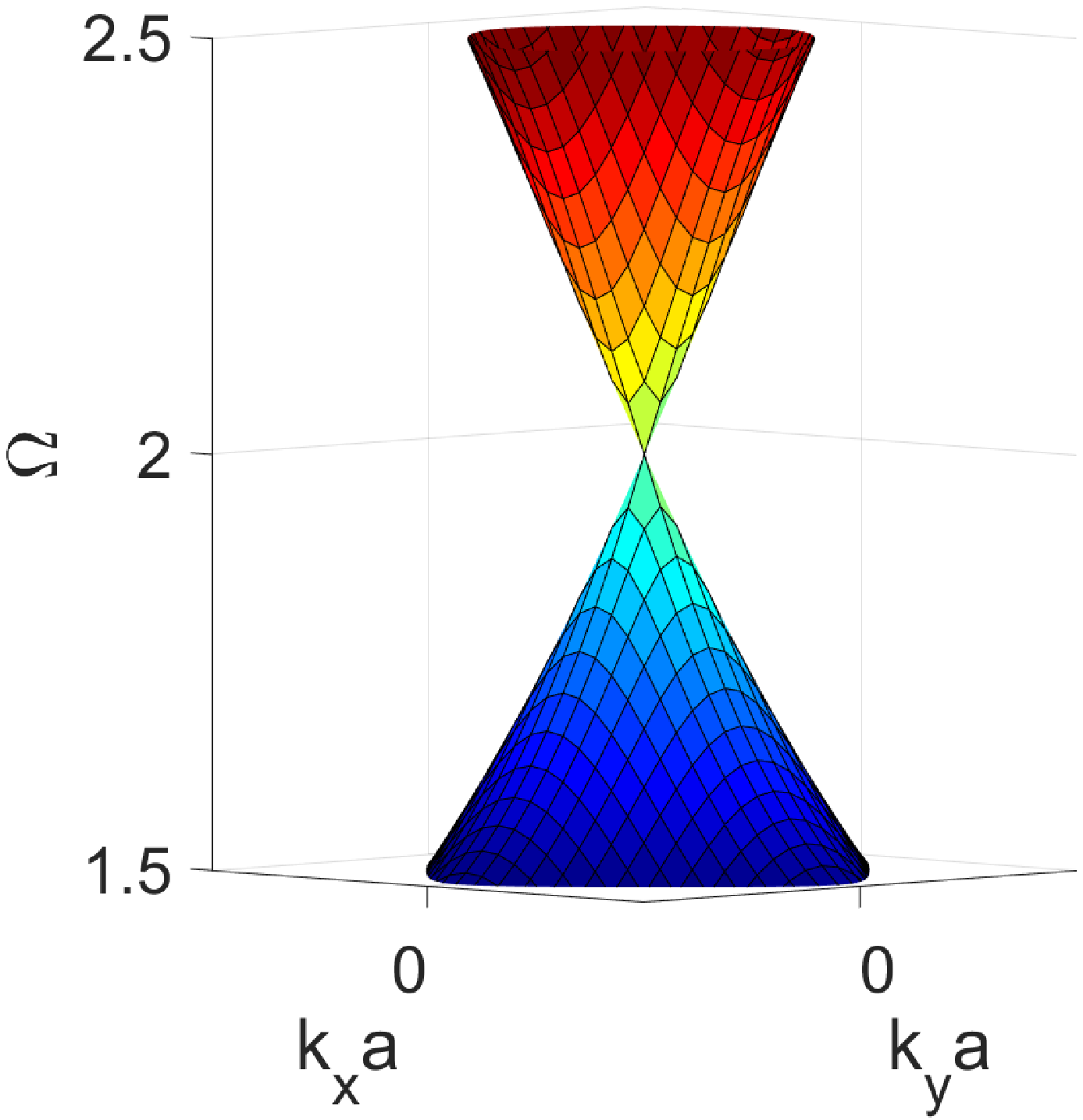}    \\
       \textbf{(c)} $P_2$ & \textbf{(d)} $P_3$ \\
     \includegraphics[height=4.5 cm, valign=c]{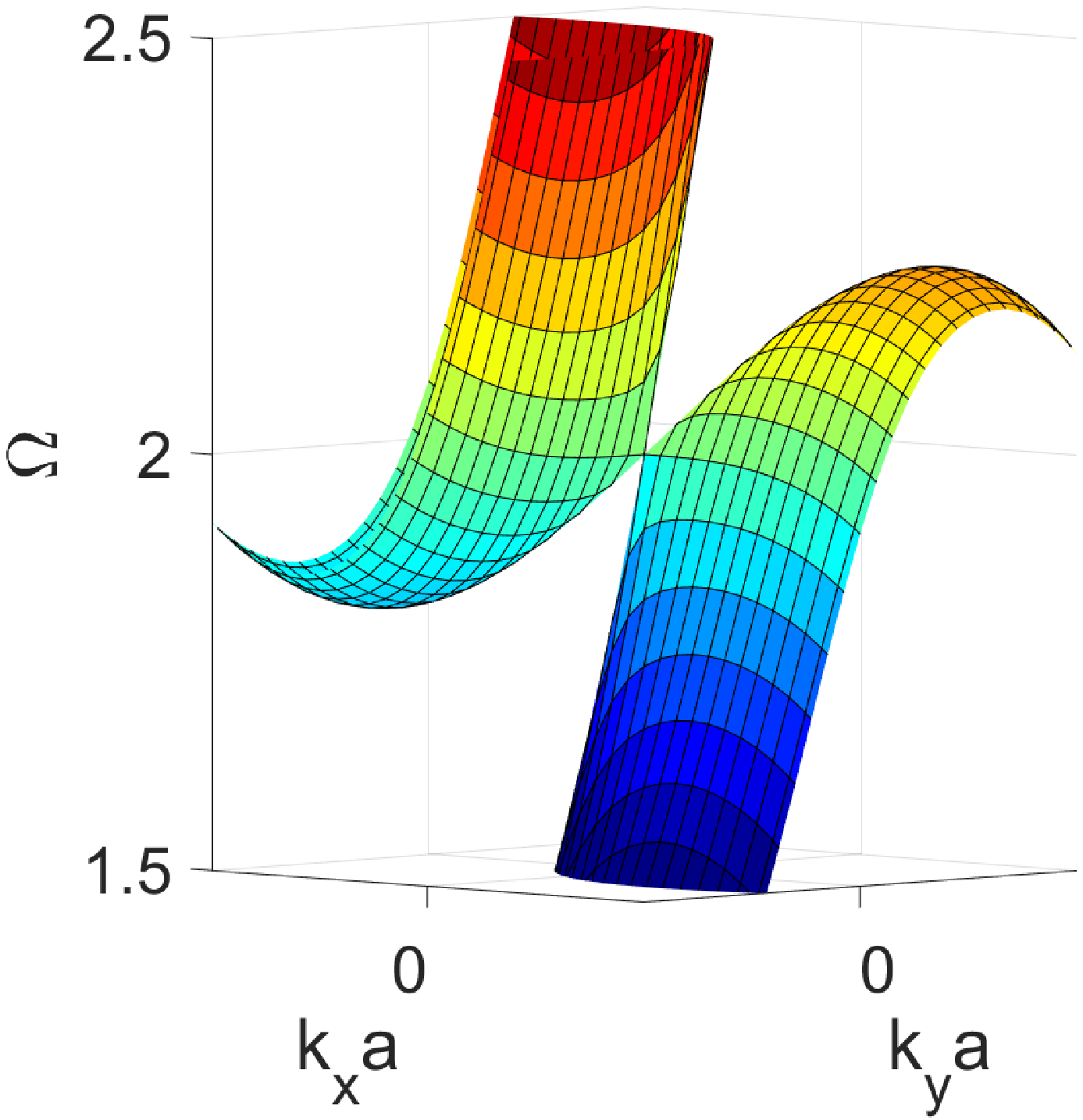}    &  \includegraphics[height=4.5 cm, valign=c]{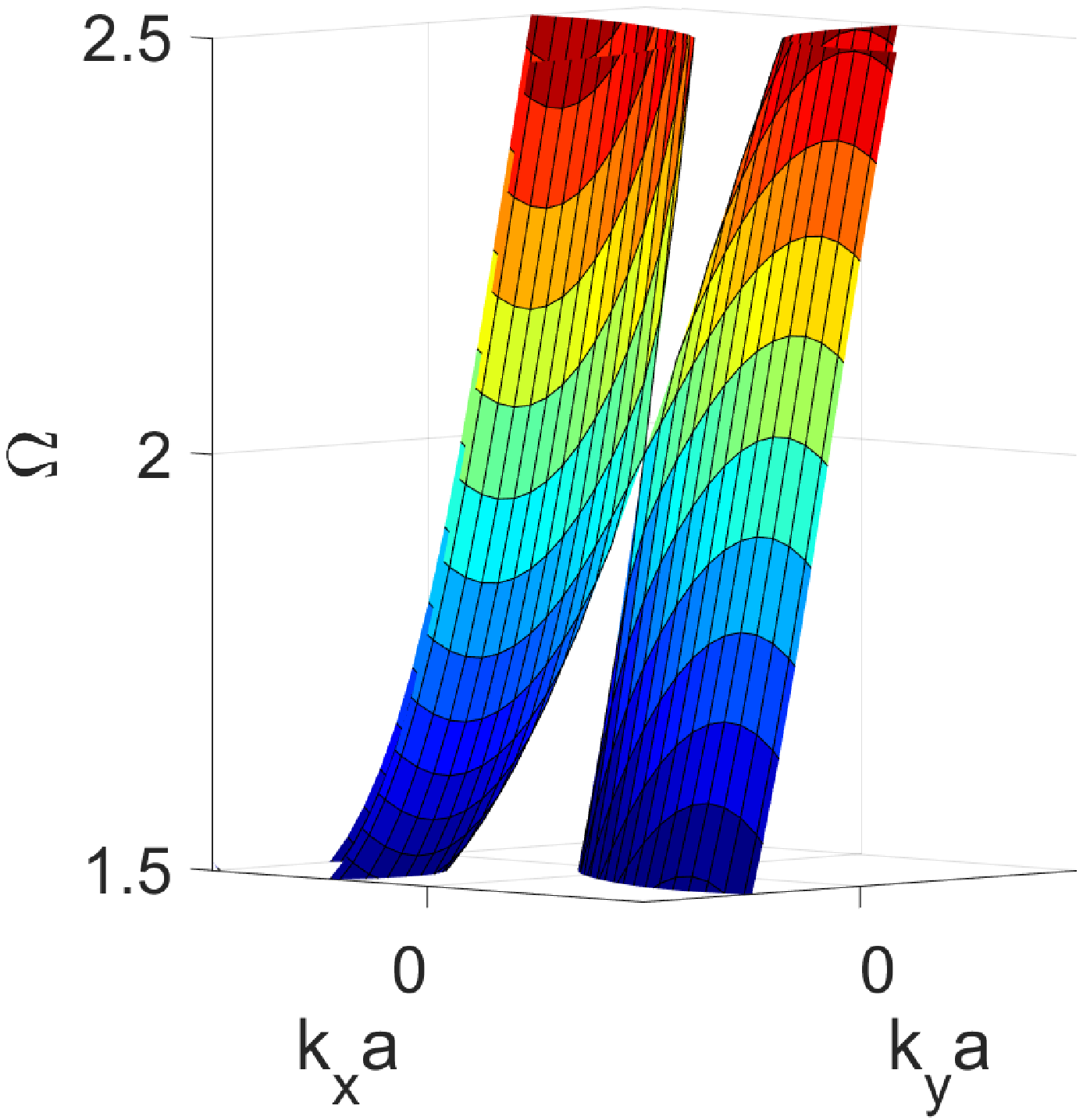}
    \end{tabular}    
    \textbf{(e)} \\
    \includegraphics[height=6.4 cm, valign=c]{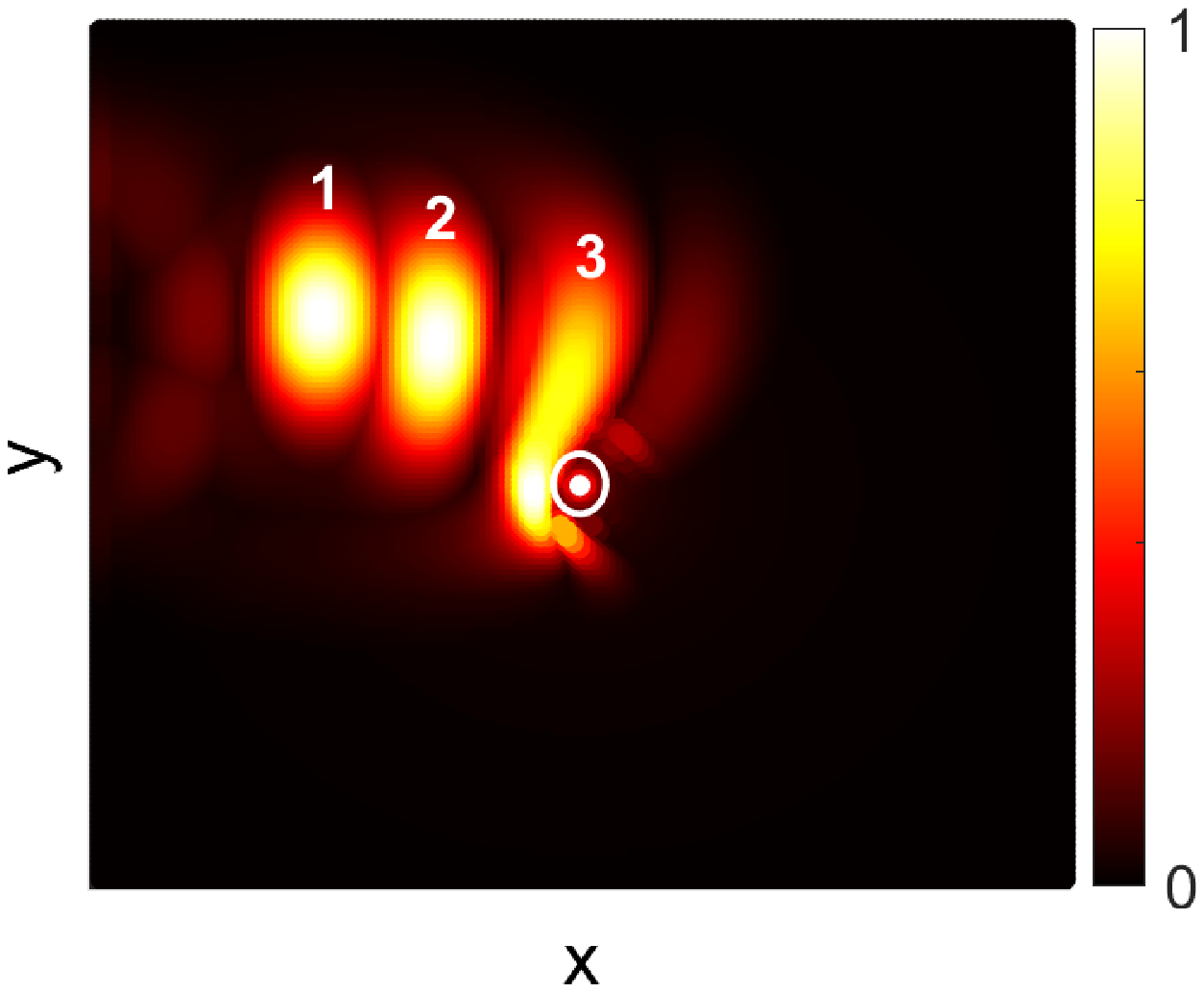} 
\end{center}
\caption{Synthetic curved spacetime validation in momentum and real space. (a) The potential $V_t(r)$. (b),(c),(d) The classical frequency spectrum corresponding to outside the hole, on the horizon and inside the hole locations, indicated by the points $P_1$, $P_2$ and $P_3$ in Fig. \ref{fig:model}(a), and respectively evolving from non-tilted, critically-tilted and over-tilted cones. (e) Time domain wavepacket propagation across the network, plotted at increasing time instances (1,2,3), and featuring the expected bending toward the black hole (white circle).}
\label{fig:performance}
\end{figure}

Next we demonstrate that the closed loop system resulting from \eqref{eq:control_eq} fully satisfies the properties of curved spacetime, both in momentum and in real space.
Due to the velocity terms, in momentum space we obtain a quadratic eigenvalue problem \cite{supplementary}. The potential is selected in the funnel form $V_t(r)=\gamma/r$, Fig. \ref{fig:performance}(a), where $\gamma$$>$0 is analogous to the black hole mass \cite{konye2022anisotropic}. The estimation of greed size in Fig. \ref{fig:model}(a) contingent upon the parameter $\gamma$. If $V_t(r)$ decays faster while crossing the horizon, the grid size is expected to be smaller compared to the scenario where $V_t(r)$ decays at a slower rate. In Fig. \ref{fig:performance}(b-d) we plot the classical frequency spectrum (zoom-in) for three representative locations across the spacetime, which are outside the hole, on the horizon and inside the hole, as respectively indicated by $P_1$, $P_2$ and $P_3$ on top of the sketch of Fig. \ref{fig:model}(a). 
As expected, outside the hole the spectrum is non-tilted, but becomes critically-tilted at the horizon, and over-tilted inside the hole.
We then numerically simulate the time domain evolution of a wavepacket, as depicted in Fig. \ref{fig:performance}(e) for three time instances. The wavepacket, which is launched far from the black hole (white circle), propagates across the network in a curved trajectory, indicating the expected bending towards the hole center. 

\begin{figure*}[tb] 
\begin{center}
\begin{tabular}{c}
\textbf{(a)} \\
\includegraphics[width=9.2cm]{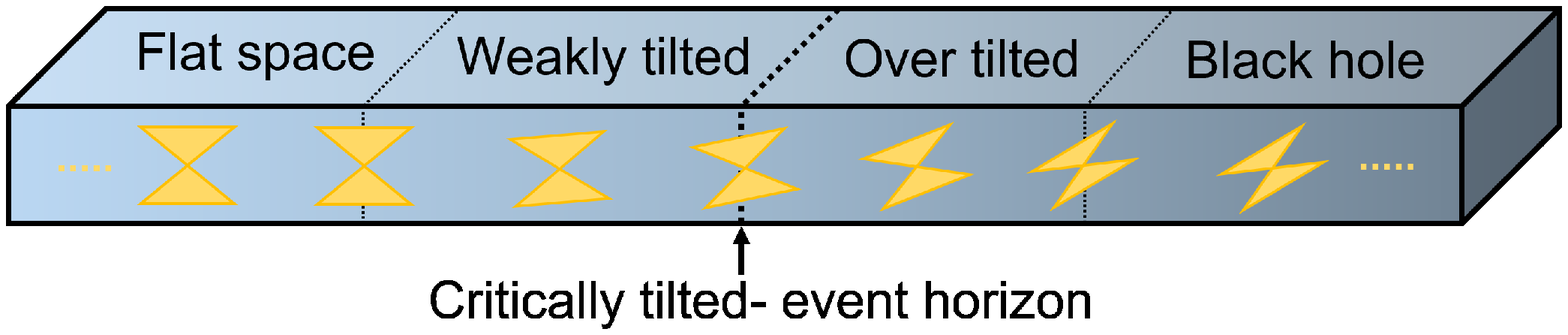}\\
\textbf{(b)}\\
\includegraphics[width=9.2cm]{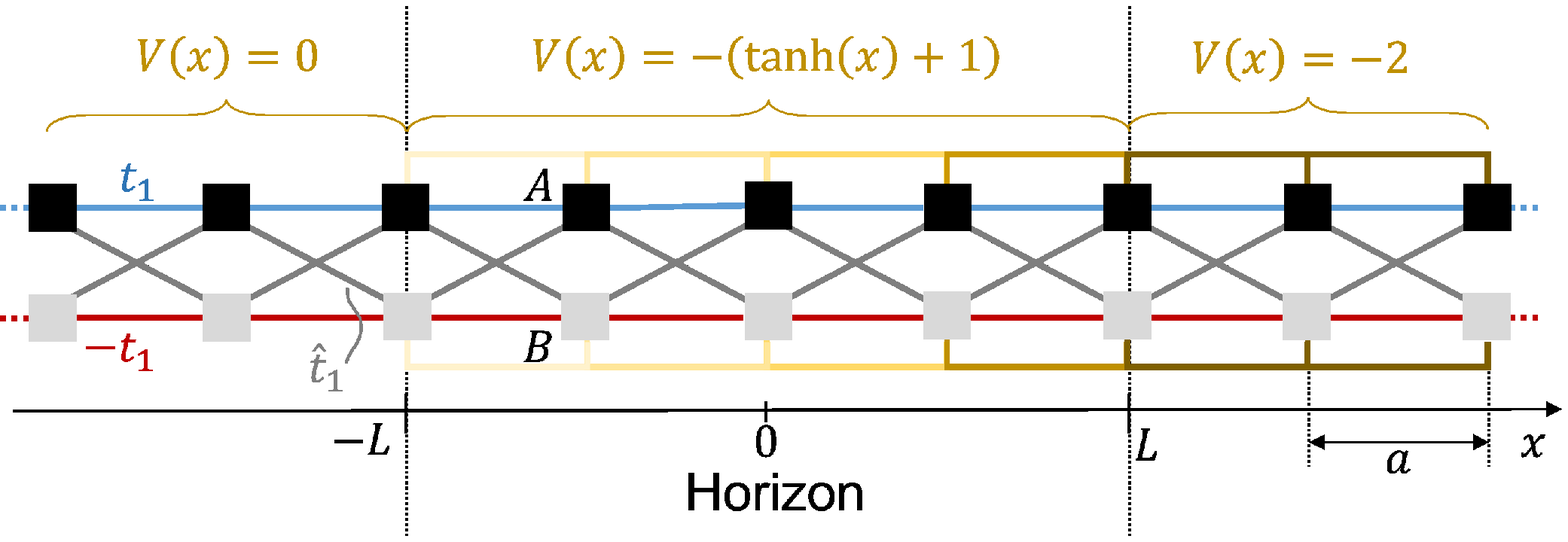} \\
\end{tabular} 
\begin{tabular}{c}
  \textbf{(g)}  \\
  \includegraphics[width=7.0cm]{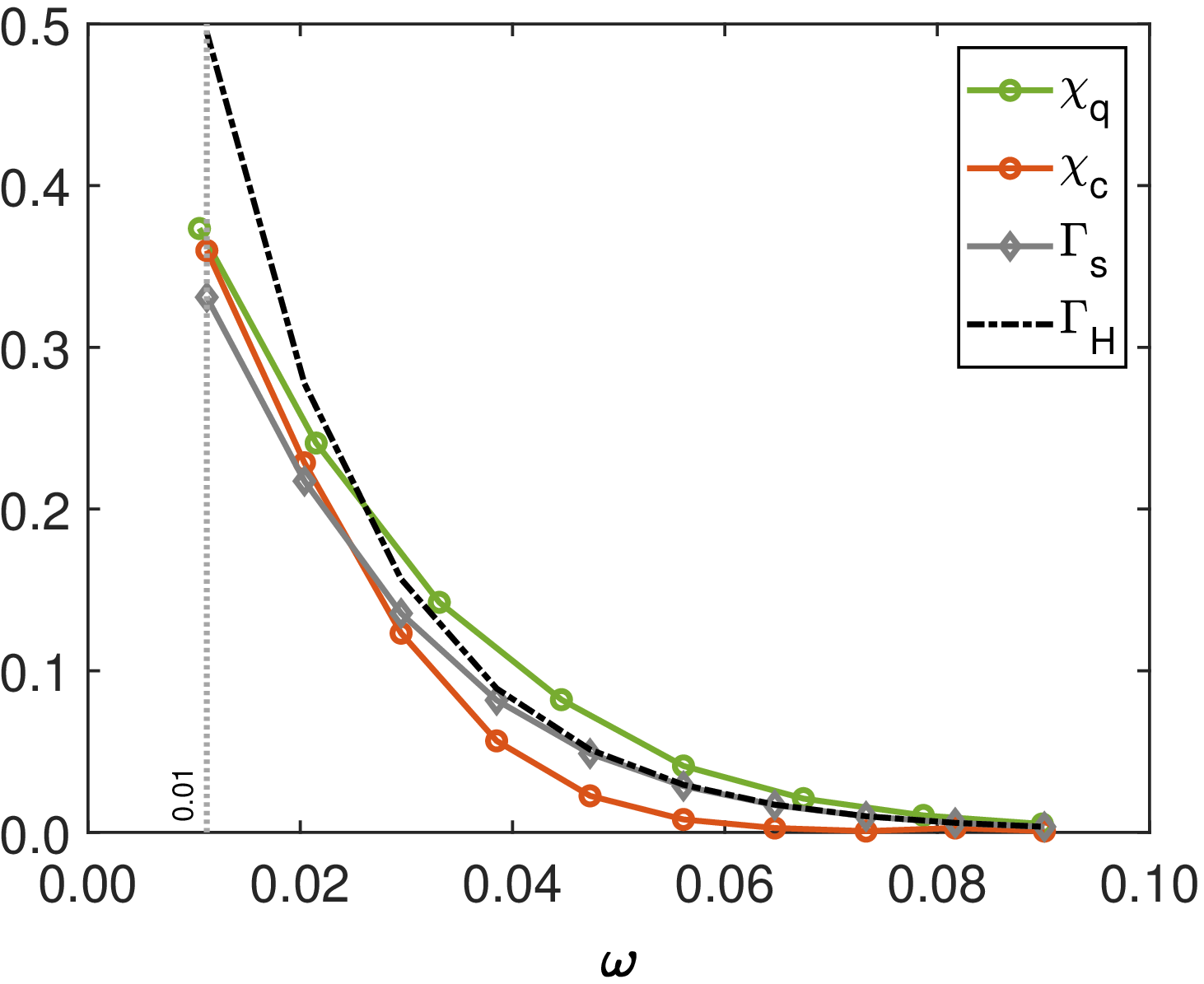}
\end{tabular}
\setlength{\tabcolsep}{-4pt}
\begin{tabular}{c c c c}
\textbf{(c)} & \textbf{(d)} & \textbf{(e)} & \textbf{(f)} \\
\includegraphics[height=\x cm, valign=c]{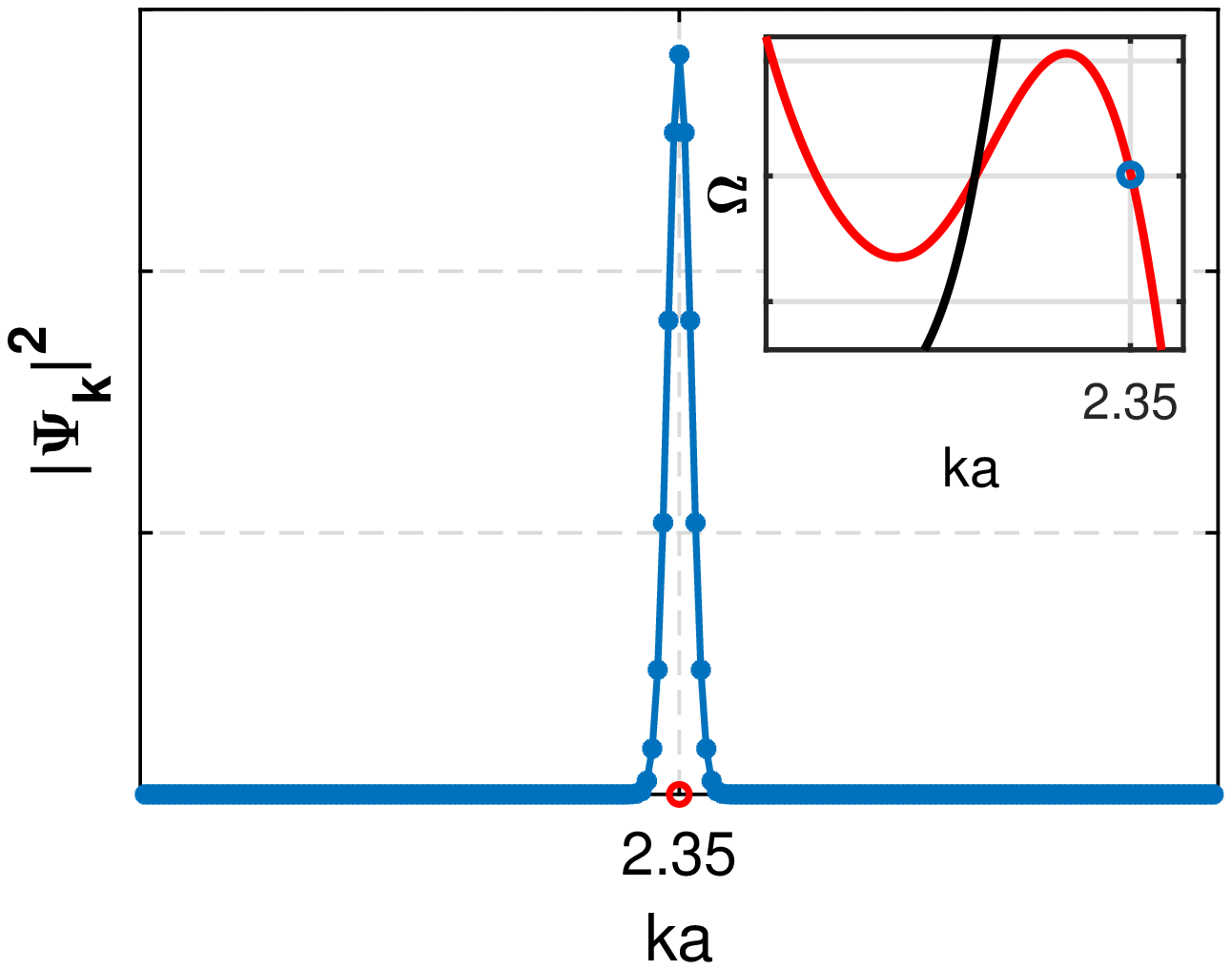} & \includegraphics[height=\x cm, valign=c]{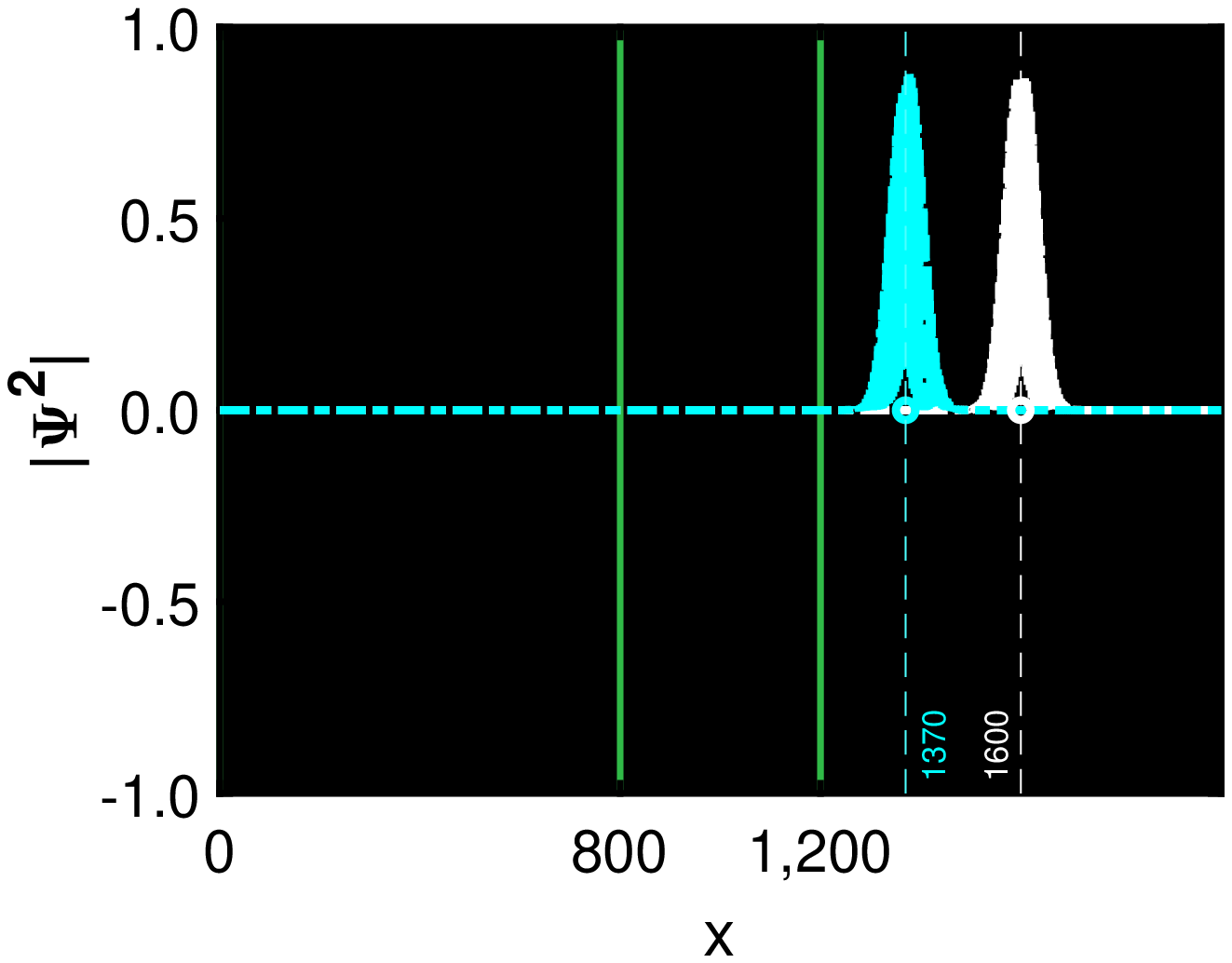} & \includegraphics[height=\x cm, valign=c]{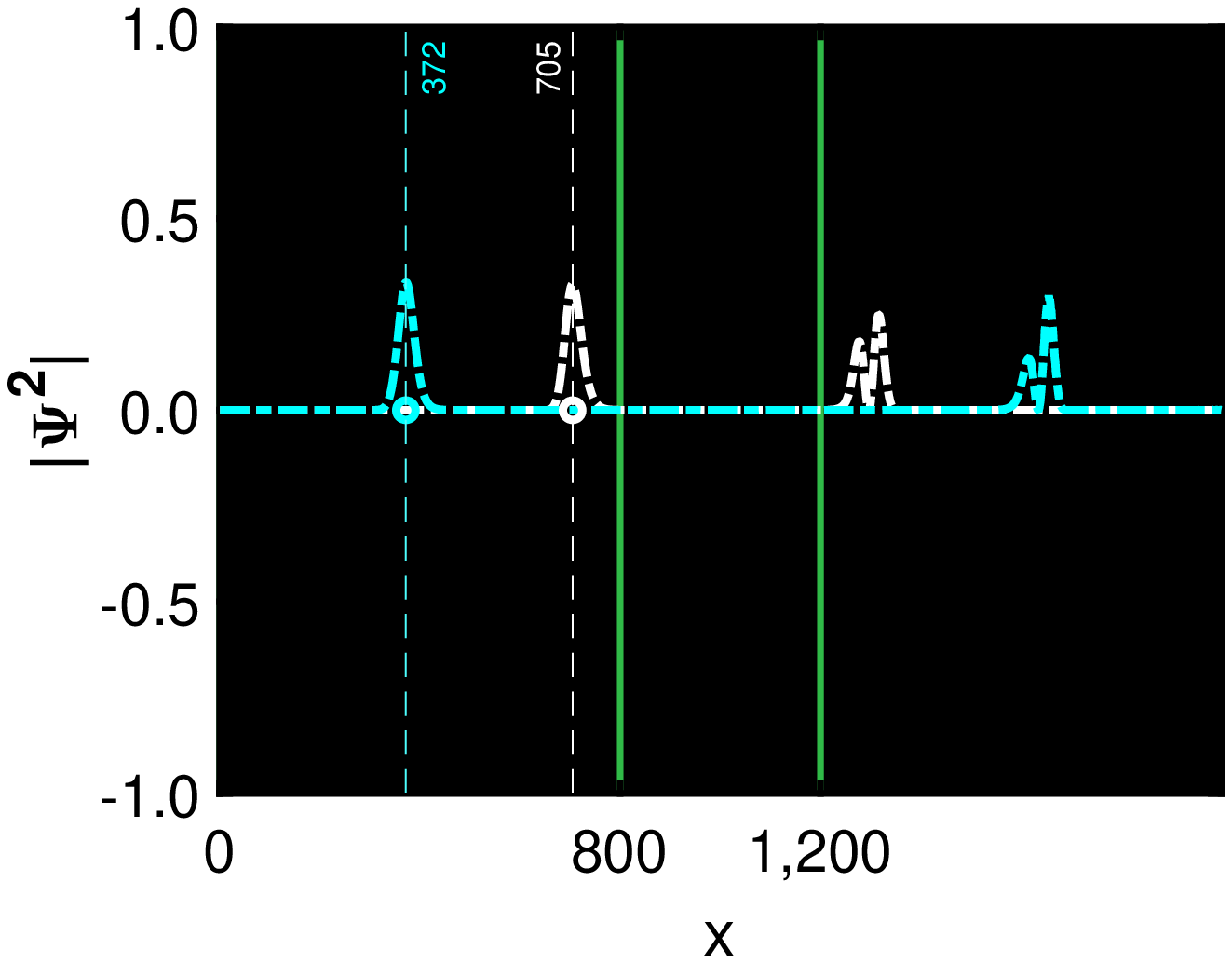} &
 \includegraphics[height=\x cm, valign=c]{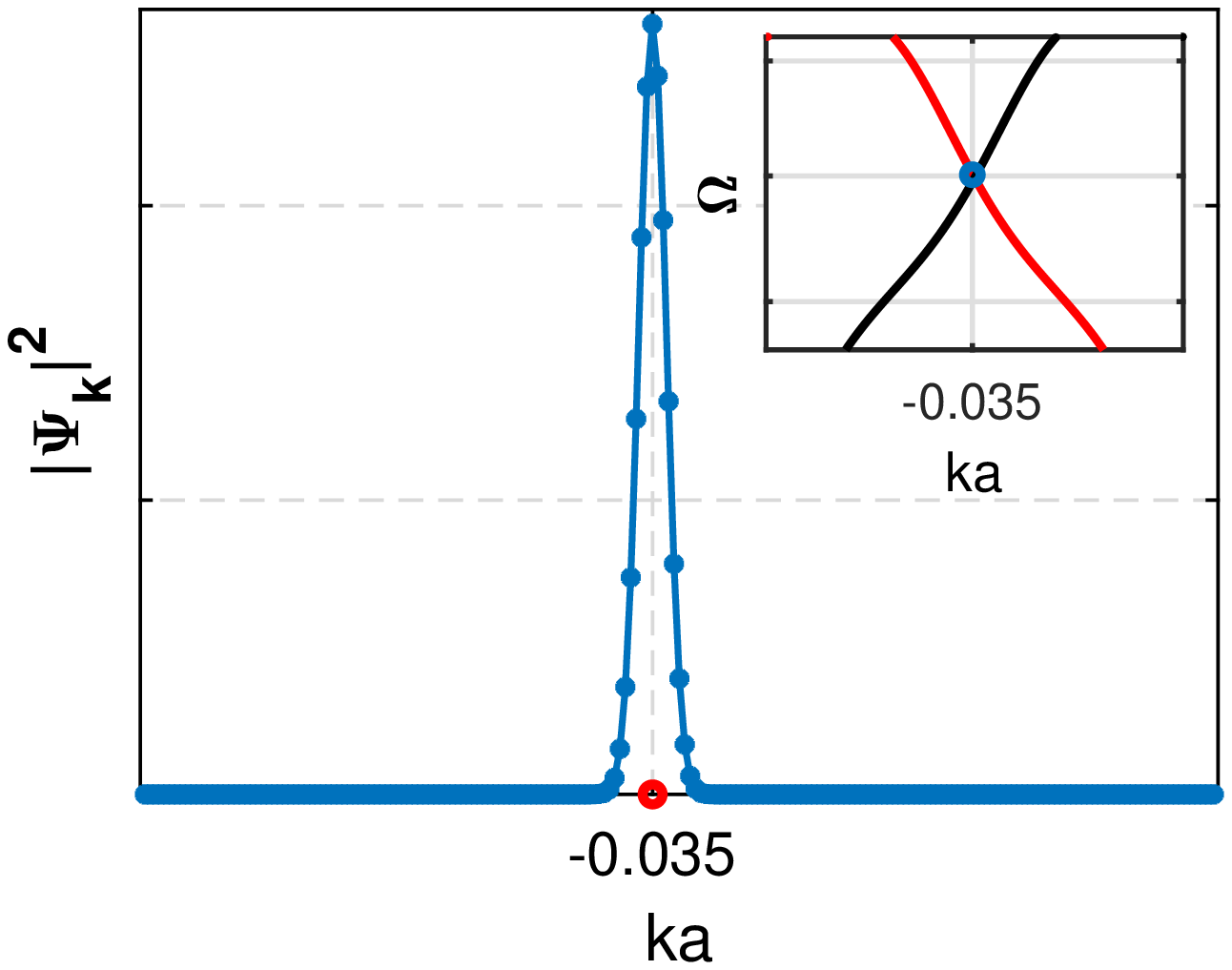}   
\end{tabular}
\caption{Artificial Hawking horizon tunneling. (a) Schematic of the WSM along the principle path. The spatially varying tilt (gold cones) represents the interface between the black hole (over-tilted cones) and flat space (zero-tilted cones). The transition occurs at the critical tilt, which represents the event horizon. (b) The relevant portion of the mechanical network of Fig. \ref{fig:model}. (c) Fourier transform of the initial wavepacket in flat space. (d),(e) Time evolution of a Gaussian wavepacket across the artificial horizon (green lines) of length $N_{0}=400$ and side sizes $N_{L},N_{R}=800$, tilt rate $\gamma_{t}=0.1$, lattice constant $a=1$, and coupling strength $t_x,t_z=1$. At time $T_{0}$ the wavepacket is launched at position $x_0=1600$. 
(d) The response (absolute value) before tunneling at $T_0$ (white) and $T_1=198$ (cyan). (e) The response after tunneling at $T_2=1174$ (white) and $T_3=1845$ (cyan). 
(f) Fourier transform of the time domain response at $T_3$.
(g) Variation with $\omega$ of the original high energy emission rate $\Gamma_H$ (black), the classical equivalence for the analytical quantum transmission probability in the condensed matter formalism $\Gamma_s$ (gray), the numerically calculated quantum decay rate $\chi_q$ (green) and the numerically calculated classical decay rate $\chi_c$ (orange) for $\gamma_{t}=0.1$.}
\label{fig:Hawking}
\end{center}    
\end{figure*}

To demonstrate the versatility of our network, we reprogram the controller in \eqref{eq:control_eq} to support the horizon tunneling phenomenon. Unlike lensing, for tunneling a one-dimensional interface of gradually tilted dispersion cones is sufficient, as illustrated in Fig. \ref{fig:Hawking}(a). The cones range from zero-tilted to over-tilted, respectively standing for flat space and the black hole. The critically-tilted cone represents the event horizon.  
We thus reprogram the controller to switch off the couplings in the $y$ direction, Fig. \ref{fig:Hawking}(b), and to generate the potential $V_t(x)=-(1+\tanh\gamma_tx)$. $\gamma_t$ determines the rate of change of $V_t(x)$ across the horizon, which is directly mapped to the gravitational field strength $g$ through $\gamma_{t}=g/c$.
This potential defines the interface $-L\leq x \leq L$ by satisfying $\tanh(L)\approx 1$. Outside the interface, i.e. at $x<-L$ and $x>L$, $V_t(x)$ takes the constant end values 0 and 2, respectively. $x=0$ is the critical tilt point indicating the artificial event horizon.  

We now validate the tunneling analogue by launching a Bloch mode modulated Gaussian wavepacket from the over-tilted region, of the Fourier transform depicted in Fig. \ref{fig:Hawking}(c), and simulating its time evolution as it tunnels through the artificial horizon to the flat space region, as depicted in Fig. \ref{fig:Hawking}(d),(e). 
The Fourier transform of the transmitted wavepacket, as predicted by energy conservation, is depicted in Fig. \ref{fig:Hawking}(f).
Both momenta are indicated by a blue circle on top of the corresponding dispersion band in the insets \cite{supplementary}.
Then, defining $\omega$ as the difference between the initial wavepacket frequency and the classical spectrum crossing point, we rewrite Eq. \eqref{eq:eq1} as $\Gamma_H=e^{-2\pi E/\gamma_t}$, which is the limit of the analytical quantum transmission probability $\Gamma_s=1/(1+e^{2\pi E/\gamma_t})$ \cite{de2021artificial,sabsovich2022hawking}, and consider the classical equivalence $\omega=E$. 
We then define the numerical decay rate for the quantum, $\chi_q$, and the classical, $\chi_c$, models, with $\Psi_m^f$ and $\Psi_m^{in}$ as the squared amplitudes of the final and initial wavepacket, obtained from time domain simulations via $\chi_{q/c}=\sum_m^{N_{L}}|\Psi_m^{f}|^2/\sum_m^{N_{R}}|\Psi_m^{in}|^2$. The dependence of $\Gamma_H$ (black), $\Gamma_s$ (gray), $\chi_q$ (green) and $\chi_c$ (orange) on $\omega$ is depicted in Fig. \ref{fig:Hawking}(g).
Remarkably, we observe that as a function of $\omega$, $\chi_c$ closely follows the profile of $\chi_q$, $\Gamma_s$ and $\Gamma_H$. This validates our system as a classical analogue of Hawking phenomena.

To conclude, we proposed a purely classical realization of artificial curved spacetime, based on WSM formalism. Our model features a two-dimensional network of mass elements the collective dynamics of which is equivalent to WSM with inhomogeneous potential and the associated varying dispersion tilt.
The resulting mechanical circuits required unstable and non-reciprocal couplings, which were created in real-time using embedded active feedback controller.
Despite the instability of the individual couplings, the control algorithm managed to stabilize the overall network.
The wide operational bandwidth of the feedback loop electrical circuitry, typically at the order of megaHertz, ensures that the latency of the mechanical system excitation, typically at the order of Hertz (Hz), is negligible. 
%

Our model produced the required space dependent tilt strength for gravitational analogies in a bulk material, thus offering enhanced experimental freedom compared to electronic Weyl semimetal \cite{yang2019topological,deng2016experimental,li2017evidence}.
Using dynamical simulations we demonstrated bending of mechanical wavepackets towards an artificial black hole in the network center, manifesting the gravitational lensing phenomenon.
By reprogramming the controller gains, we mimicked horizon tunneling on the same platform. 
The attenuation rate of a tunneled wavepacket matched well the transmission probability of the quantum system, as well as the emission rate of the original black hole.
The reprogrammable nature of our platform enables to test, e.g., the effect of black holes of different sizes on lensing and tunneling, and on other high energy phenomena related to curved space time.

\textit{The authors are grateful to Steven Cummer, John Smith, Martin Wegener, Romain Fleury, Badreddine Assouar, Bogdan Popa, Jensen Li, Vincenzo Vitelli and Chen Shen for insightful discussions. A special thanks goes to Daniel Sabsovich for important comments at the beginning of this project.}

\bibliography{paper}

\end{document}